\documentclass{ws-p8-50x6-00}
\def\UT{University of Tokyo}
\def\hongo{7-3-1 Hongo, Bunkyo-ku, Tokyo 113-0033, Japan}
\def\gagg{g_{a\gamma\gamma}}

\begin{document}

\title{Recent results from the Tokyo axion helioscope experiment}

\author{Y. Inoue$^{\lowercase{ae}}$,
  T.~Namba$^{\lowercase{b}}$, S.~Moriyama$^{\lowercase{c}}$,
  M.~Minowa$^{\lowercase{be}}$,
  Y.~Takasu$^{\lowercase{b}}$, T.~Horiuchi$^{\lowercase{b}}$,
  and~A.~Yamamoto$^{\lowercase{d}}$}
\address{
  ${}^a$International Center for Elementary Particle Physics,
  \UT,\\ \hongo\\
  ${}^b$Department of Physics, School of Science, \UT,\\ \hongo\\
  ${}^c$Kamioka Observatory, Institute for Cosmic Ray Research, \UT,\\
  Kamioka-cho, Yoshiki-gun, Gifu 506-1205, Japan\\
  ${}^d$High Energy Accelerator Research Organization (KEK),\\
  1-1 Oho, Tsukuba, Ibaraki 305-0801, Japan\\
  ${}^e$Research Center for the Early Universe (RESCEU), \UT,\\
  \hongo}

\maketitle

\abstracts{
We have searched for axions which could have been produced in the
solar core using an axion helioscope
which is equipped with
a 2.3\,m-long 4\,T superconducting magnet,
PIN-photodiode x-ray detectors,
and a telescope mount mechanism to track the sun.
A gas container to hold dispersion-matching gas
has been developed and
a mass region up to $m_a=0.26\rm\,eV$
was newly explored.
Preliminary analysis sets a limit on axion-photon
coupling constant to be
$\gagg<6.4\sim9.6\times10^{-10}\rm GeV^{-1}$
for the axion mass of
$0.05<m_a<0.26\rm\,eV$
at 95\% confidence level from the absence of the axion signal.
This is more stringent than the limit inferred from
the solar age consideration
and also more stringent than the recent helioseismological bound.}

\section{Introduction}
The axion which is renowned as
a dark matter candidate
is the Nambu-Goldston boson of the Peccei-Quinn symmetry
introduced to solve
the strong $CP$ problem
in the strong interaction theory.~\cite{axion-th,axion-rev}
The expected behavior of an axion is characterized
mostly by its mass, $m_a$.
If $10^{-5}<m_a<10^{-3}\rm\,eV$,\footnote{The lower bound can be
even lower depending on the scenario.}
the axion can be copiously produced in the early universe
so that they can close the universe.
Another astrophysically interesting mass region is
at around one to a few eV.
Such axions can be allowed for some hardonic axion models
by other astrophysical or cosmological constraints.
In this region,
the sun can be a powerful source of axions
and the so-called `axion helioscope' technique
may enable us to detect such axions directly.~\cite{Sikivie,Bibber}

\begin{figure}[t]
  \begin{center}
    \epsfbox{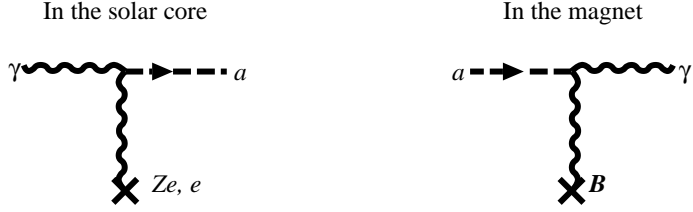}
  \end{center}
  \caption{The solar axions produced via the Primakoff process
    in the solar core are, then, reconverted into x rays
    via the reverse process.}
  \label{fig:principle}
\end{figure}
The principle of the axion helioscope is illustrated
in Fig.~\ref{fig:principle}.
In the solar core, axions can be abundantly converted
from black body radiation photons through the Primakoff process.
Then, they are coherently reconverted into x~rays in a strong magnetic
field in a laboratory.
Their average energy is 4.2\,keV
reflecting the original black body radiation.
The conversion rate is given by:
\begin{equation}
  P_{a\to\gamma}
  ={\gagg^2\over4}\left|\int_0^L Be^{iqz}dz\right|^2
\end{equation}
where
$\gagg$ is the axion-photon coupling constant,
$z$ is the coordinate along the incident solar axion,
$B$ is the strength of the magnetic field,
$L$ is the length along the $z$-axis,
and
$q=|(m_\gamma^2-m_a^2)/2E|$ is the momentum transfer by
the virtual photon.
Here, $m_\gamma$ is the effective mass of the photon
which is of course zero in vacuum.

We have constructed a helioscope with a dedicated superconducting
magnet,
and the first measurement~\cite{sumico97} was performed
during from 26th to 31st December 1997.
From the absence of the axion signal,
an upper limit to the axion-photon coupling was set to be
$\gagg<6.0\times10^{-10}\rm GeV^{-1}$ (95\% CL)
for $m_a<0.03\rm\,eV$.
For heavier axions,
momentum transfer, $q$, becomes not negligible,
thus
the sensitivity of the detector is lost.

Coherence can, however, be restored
by filling the conversion
region with buffer gas
since a photon of x-ray region has an effective mass
in a medium.
For light gas, such as hydrogen or helium,
it is simply written as:
\begin{equation}
  m_\gamma=\sqrt{4\pi\alpha N_e\over m_e}.
\end{equation}
We adopted cold helium gas as the dispersion-matching medium
and scanned the mass region up to 0.26\,eV.
In this paper, we will present a preliminary result of
this new measurement.

\section{Experimental apparatus}

\begin{figure}
  \begin{center}
    \epsfbox{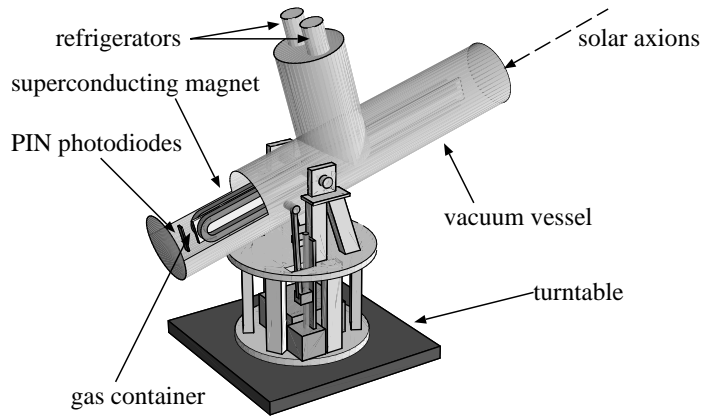}
  \end{center}
  \caption{The schematic view of the axion helioscope.}
  \label{fig:sumico}
\end{figure}

The schematic figure of the helioscope is shown
in Fig.~\ref{fig:sumico}.
The detector consists of
a superconducting magnet, x-ray detectors, a gas container,
and an altazimuth mounting.

The superconducting magnet was
so designed as to be easy to swing.
It consists of two 2.3-m long
race-track shaped coils run parallel
with a 20-mm wide gap between them.
The magnetic field in the gap is 4\,T
perpendicular to the helioscope axis.
The coils are kept at 5--6\,K during operation.
They are directly cooled with
two Gifford-McMahon refrigerators and
no cryogen is needed.
The magnet has also a persistent current switch.
After excitation,
it is switched into persistent current mode
and the current leads are taken away.

Sixteen PIN photodiodes, Hamamatsu Photonics S3590-06,
are used as the x-ray detectors.
The chip size is $11\times11\times0.5\rm\,mm^3$.
Each chips is mounted on a Kapton film
bonded to an Invar plate with cryogenic compatible adhesive.
The x-ray detectors
are mounted in a radiation shielding box made of oxygen-free
high conductivity copper (OFHC Cu)
which is operated at about 60\,K.
The copper shield is surrounded by a lead shield
which is at room temperature.

The output from each photodiodes is fed
to a charge sensitive preamplifier whose first-stage
FET is at the cryogenic stage near the photodiode chip
and the preamprefier outputs are digitized using
flash analog-to-digital convertors (FADC's), REPIC RPC-081's.
We performed numerical pulse shaping to the raw waveform
using the Wiener filter.
The energy of an x ray is given by the peak height of
a wave after shaping.
Each detectors was calibrated by 6-keV manganese x~ray
from a $^{55}$Fe source which can be exposed and
hided freely during the measurement.
The energy resolutions for 6-keV photon was
0.8--1\,keV (FWHM).

The new device introduced this time is the gas container.
We adopted cold helium gas as the dispersion-matching medium.
Light gas is preferred since it minimizes x-ray absorption
by the gas.
Helium has another virtue that it remains at gas state
at 5\,K, the same temperature as the coil.
At 5\,K, the gas pressure corresponding to
our ultimate goal, $m_a=2.6\rm\,eV$,
reaches only 0.13\,MPa.

The container body is made of four stainless steel square pipes
welded to each other.
The entire container body is wrapped with 5N high purity
aluminium sheet to achieve high uniformity of temperature.
At the end of the container, gas is separated from vacuum
with an x-ray window
which is transparent to x ray above 2\,keV
and can hold gas up to 0.3\,MPa at liquid helium temperature.

The whole helioscope is constructed in a vacuum vessel
and is mounted on an altazimuth mount.
Its trackable altitude ranges from $-28^\circ$ to $+28^\circ$
and almost any azimuthal direction is trackable.
This view corresponds to
about 50\% of duty cycle for the sun measurement in Tokyo.
The other half of a day we measure the background.
This helioscope mount is driven by two AC servo motors
controlled by a personal computer (PC) through CAMAC bus.
The PC regularly monitors two precision rotary encoders
through CAMAC bus and forms a feedback control loop.
The U.S. Naval Observatory Vector Astronomy Subroutines (NOVAS)~\cite{novas}
was used to calculate the sun position.
The directional origin of the helioscope was measured
using a theodolite.
The absolute azimuth is determined from the observed
direction of Polaris
and the absolute altitude is determined from a spirit level.

\section{Measurement and Analysis}
During from 29th July to 1st September 2000,
a new measurement with buffer gas was performed 
for ten photon mass settings
to scan up to 0.26 eV.

\begin{figure}[t]
  \begin{center}
    \epsfbox{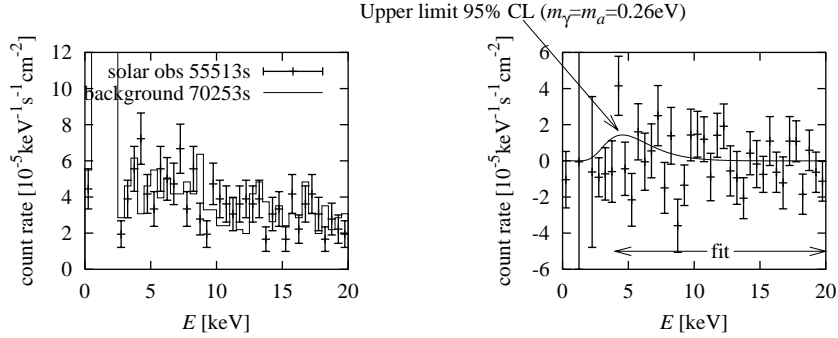}
  \end{center}
  \caption{The left figure shows the energy spectrum
    of the solar observation (error bars) and
    the background spectrum (solid line) when
    the gas density was tuned to $m_\gamma=0.26\rm\,eV$.
    The right figure shows the net energy spectrum of the left
    where the background is subtracted from the solar observation.}
  \label{fig:spec}
\end{figure}

In Fig.~\ref{fig:spec},
one of the energy spectra of the solar observation
is shown together with the background spectrum.
We searched for expected axion signals
which scales with $\gagg^4$
for various $m_a$
in these spectra.
The smooth curve in the figure represents an example for
the expected axion signal
for $m_a=m_\gamma=0.26\rm\,eV$ and $\gagg=7\times10^{-10}\rm GeV^{-1}$.

In the analysis,
the ten measurements with the ten different gas densities
are combined together by using the total $\chi^2$.
The energy region 4--20\,keV was used for fitting.
Since no significant excess was seen for any $m_a$,
we set upper limit on $\gagg$ at 95\% confidence level
using the Bayesian method.

\begin{figure}[t]
  \hbox{\vtop{\vskip0pt\hbox{%
        \epsfbox{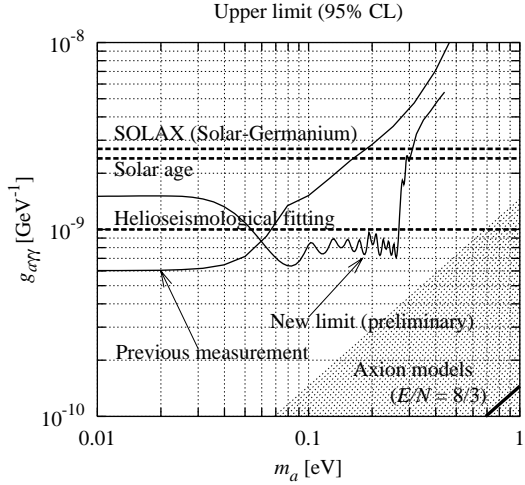}}}%
    \hskip13pt
    \vtop{\hsize=4cm
      \caption{The exclusion plot on $\gagg$ to $m_a$
        at 95\% confidence level is plotted
        where some other bounds are plotted together.
        Our previous limit and the new limit is plotted in solid lines.
        Dashed lines are
        the limit by SOLAX experiment,
        the limit inferred from the solar age consideration,
        and the recent helioseismological bound.
        The hatched area is the preferred axion models.
        The thick line corresponds to the case when
        a simple GUT is assumed. 
        }}}
  \label{fig:exclusion}
\end{figure}

In Fig.~\ref{fig:exclusion},
upper limits on $\gagg$ are plotted as a function of $m_a$.
The previous limit and the new preliminary limit are plotted together.
There are also some other bounds plotted in the same figure.
The SOLAX~\cite{solax} is a solar axion experiment
which exploits the coherent conversion
on the crystalline planes in a germanium detector.
The limit $\gagg<2.3\times10^{-9}\rm GeV^{-1}$ is
the solar limit inferred from the solar age consideration.
The limit $\gagg<1\times10^{-9}\rm GeV^{-1}$
is a new solar limit recently reported.~\cite{helioseismology}
Above this line,
standard solar models with energy losses by solar axions
cannot fit to the helioseismological sound-speed profile.

\section{Conclusion}
We have developed a gas container and
introduced cold helium gas
as the dispersion-matching medium
in the $4{\rm\,T}\times2.3\rm\,m$ magnetic field
of our axion helioscope.
The axion mass up to 0.26eV has been newly scanned.
But no evidence for solar axions was seen.
New preliminary upper limit on $\gagg$
which ranges 6.4--$9.6\times10^{-10}\rm GeV^{-1}$
was set for $0.05<m_a<0.26\rm\,eV$,
which is far more stringent than the solar existence limit
and is also more stringent than the recent helioseismological bound.
This experiment is currently the only one existing
experiment which has enough sensitivity to detect such
solar axion that do not violate the solar model itself.

\section*{Acknowledgments}
This research is supported by the Grant-in-Aid for COE research
by the Japanese Ministry of Education, Science, Sports and Culture,
and also by the Matsuo Foundation.

\end{document}